\documentclass[pra,twocolumn,superscriptaddress,amsmath,amssymb]{revtex4-1}
\usepackage{amsmath}
\usepackage{amsfonts}
\usepackage{amssymb}
\usepackage{graphicx}
\usepackage{color}
\usepackage{txfonts}
\usepackage{float}
\usepackage[titletoc]{appendix}
\usepackage[colorlinks={true}]{hyperref}
\usepackage{textcase}
\hypersetup{citecolor={blue}, filecolor={blue}, linkcolor={blue}, urlcolor={blue}}

\begin{document}

\title{\textbf{Precise quantum control of unidirectional field-free molecular orientation}
}
\author{Qian-Qian Hong}
\affiliation{Hunan Key Laboratory of Super-Microstructure and Ultrafast Process, School of Physics, Central South University, Changsha 410083, China}
\author{Zhe-Jun Zhang}
\affiliation{Hunan Key Laboratory of Super-Microstructure and Ultrafast Process, School of Physics, Central South University, Changsha 410083, China}
\author{Chuan-Cun Shu}
\email{cc.shu@csu.edu.cn}
\affiliation{Hunan Key Laboratory of Super-Microstructure and Ultrafast Process, School of Physics, Central South University, Changsha 410083, China}
\author{Jun He}
\affiliation{Hunan Key Laboratory of Super-Microstructure and Ultrafast Process, School of Physics, Central South University, Changsha 410083, China}
\author{Daoyi Dong}
\affiliation{Australian Artificial Intelligence Institute, Faculty of Engineering and Information Technology, University of Technology Sydney, NSW 2007, Australia}
\author{Dajun Ding}
\email{dajund@jlu.edu.cn}
\affiliation{Institute of Atomic and Molecular Physics, Jilin Provincial Key Laboratory of Applied Atomic and Molecular Spectroscopy, Jilin University, Changchun 130012, China}

\begin{abstract}
The capability to control molecular rotation for field-free orientation, which arranges molecules in specific spatial directions without external fields, is crucial in physics, chemistry, and quantum information science. However, conventional methods typically lead to transient orientations characterized by periodic directional reversals and necessitate the generation of coherent superpositions across a broad spectrum of rotational states of ultracold molecules.  In this work, we develop a theoretical framework for achieving unidirectional field-free orientation by selectively manipulating two specific rotational states of symmetric top molecules. By leveraging the interplay between coherent superpositions and the precise selection of initial states, we demonstrate that both the maximum achievable orientation and its direction can be effectively controlled. To attain the desired two-state orientation, we present a quantum control strategy that utilizes a single control pulse, significantly simplifying the complexities of conventional multistate or multipulse schemes. Numerical simulations validate the effectiveness and feasibility of this approach for methyl iodide (CH$_3$I) molecules, even when accounting for molecular centrifugal distortion.The results highlight the critical roles of initial-state selection and quantum coherence in achieving long-lasting, high unidirectional molecular orientation, opening new directions in stereochemistry, precision spectroscopy, and quantum computing.
\end{abstract}
\maketitle
\section{Introduction}
Manipulating the rotational degrees of freedom of molecules, particularly achieving molecular alignment and orientation \cite{2003RMP_seideman,2019RMP_Sugny,2023PCCP_shu}, is of great interest in physics \cite{2018PRL_James,2019NC_Karamatskos,2022NC_Mullins,2025SA_Guo,PhysRevLett.133.263601}, chemistry \cite{2019NC_Fang,2022PRL_Li,2024NC_Tanaka,2024NC_Willitsch}, and quantum information science \cite{2002PRL_DeMille,2020PRX_Albert,2024PRL_Ye,2024NP_Cornish}. Molecular alignment refers to the spatial organization of molecular axes, whereas orientation aligns molecules in a specific spatial direction. To evaluate the degree of molecular orientation, the expectation value of the orientation operator, \(\langle\cos\theta\rangle\), serves as a reliable metric, where \(\theta\) denotes the angle between the molecular axis and a designated reference axis. A perfect orientation, characterized by \(|\langle\cos\theta\rangle|=1\), indicates that all molecules are precisely aligned with the reference axis, whereas a value of \(\langle\cos\theta\rangle=0\) corresponds to randomly oriented molecules.
 In addition to using “brute force”, such as direct current (dc) fields \cite{1991Nature_Friedrich}, to obtain spatial control of molecular
 orientation, considerable theoretical and experimental efforts have been dedicated over the past three decades to achieving field-free molecular orientation \cite{2001PRL_Niels}, which refers to the directional preference of a molecular axis in the absence of external fields. The underlying physics used in conventional methods to achieve this phenomenon generally involves either resonant interactions of tailored terahertz or microwave pulses with the molecular permanent dipole moment \cite{2010JCP_Shu,2011PRL_Fleischer,2013PRA_Itatani,2020PRA_Shu,2021PRR_Averbukh}, or nonresonant interactions of intense ultrashort laser pulses with the molecular polarizability and hyperpolarizability \cite{,2009PRL_Kling,2010PRL_Sakai,2012PRL_Spanner,2014PRL_Kraus,2014PRL_Kling,2014PRA_Kumarappan,2018NC_Wu,2024PRA_Shu}, as well as their hybrids \cite{2011PRA_Itatani,2013PRA_Shu,2014PRL_Jones,2016PRL_Fleischer,2021NJP_Friedrich,2023JCP_Ran}. These interactions result in a coherent superposition of rotational states, leading to the field-free orientation of molecules, which periodically change their spatial direction.  To date, conventional methods usually require excitation across many rotational states, resulting in transient and short-lasting orientation. This implies that achieving and sustaining a high and long-lasting molecular orientation remains a significant challenge \cite{2005PRL_Sugny,2020PRL_Xu,2024PRR_IlyaSh}.  \\ \indent
 Recent breakthroughs in theory and experimentation have demonstrated that optimal control over superpositions of a limited number of rotational states can achieve a high degree of orientation. By optimizing resonant pulse sequences or intense nonresonant ultrashort laser pulses, theoretical studies have demonstrated that coherent superpositions tailored across the lowest two, three, four, and five rotational states can achieve maximum degrees of orientation at \( |\langle\cos\theta\rangle|_{\rm{max}} = 0.577 \), \( 0.775 \), \( 0.861 \), and \( 0.906 \), respectively, for ultracold diatomic and linear symmetric polar molecules \cite{2020PRA_Niels,2021PRA_shu,2023PRL_shu,2023JPA_Fan,2023PRA_Henriksen,2025PRA_Fan}. Through precise control of 15 microwave pulses, each with optimized amplitudes, phases, and delays, our recent theoretical work has shown that generating a desired superposition of the lowest 16 rotational states can lead to an almost perfect orientation, with \(|\langle \cos\theta \rangle|_{\rm{max}} > 0.99\) for ultracold diatomic polar molecules \cite{2025PRR_Shu}. Experimentally, optimal two-state orientation has been obtained in ultracold carbonyl sulfide (OCS) molecules by shaping a strong, non-resonant ultrashort laser pulse in conjunction with a weak dc field, yielding \(|\langle\cos\theta\rangle|_{\text{max}} \approx 0.6\) \cite{2015PRL_Trippel}.  Currently, further enhancements to incorporate additional rotational states beyond the two-state model face technical challenges related to designing optimized ultrashort pulses or multi-tone pulses. These difficulties hinder the experimental implementation of the theoretically proposed methods for optimizing the superposition of multiple rotational states.\\ \indent
To tackle the challenges of experimental implementation, it is crucial to develop a method that does not rely on complex pulse sequences, high-intensity ultrashort pulses, and ultralow temperatures. In this work, we propose and numerically validate a theoretical framework for generating and maximizing unidirectional field-free orientation in symmetric top molecules. Mathematically, unidirectional orientation indicates that the time-dependent degree of molecular orientation, i.e.,  $\langle\cos\theta\rangle(t)$, consistently maintains positive or negative values, which is advantageous for maintaining long-lasting orientation. In practice, such orientation could in principle be observed in symmetric top molecules using inhomogeneous electric fields \cite{1988Nature_Stolte, 2001Nature_Meijer, 2004S_Maurice}, which selectively transmit molecules in a specific initial state.  However, achieving a high degree of orientation with this technique remains challenging because it requires selecting an extreme rotational state. We demonstrate that unidirectional orientation can be significantly enhanced by manipulating the coherence of a selected initial rotational state in conjunction with its adjacent rotational state. To facilitate this process, we derive a two-state pulse-area theorem to assist in designing a single control pulse optimized for the superposition of two selected rotational states. Initial state selection using hexapole techniques has been demonstrated for CH$_3$Br and CH$_3$I \cite{2015PRA_Ding,2017JPCA_Ding}. While isolating a pure high-$J_0$ state is challenging, our work provides a theoretical upper bound for enhancing unidirectional field-free orientation in symmetric top molecules, with implications for  precision spectroscopy \cite{2012PRL_Kraus, 2020PRR_Lew}, chemical reaction control \cite{2015NC_Feringa, 2025AM_Kong}, and fundamental physics \cite{2019NC_Grill, PhysRevLett.133.113203}. \\ \indent
The remainder of this paper is organized as follows. Section \ref{TM} presents the theoretical method for generating unidirectional field-free orientation in symmetric top molecules. Section \ref{RD} details our numerical simulations
and analysis on methyl iodide molecules. We summarize our findings and discuss future directions in Sec. \ref{CO}.
\begin{figure*}[htbp]
\centering
\resizebox{1\textwidth}{!}{%
\includegraphics{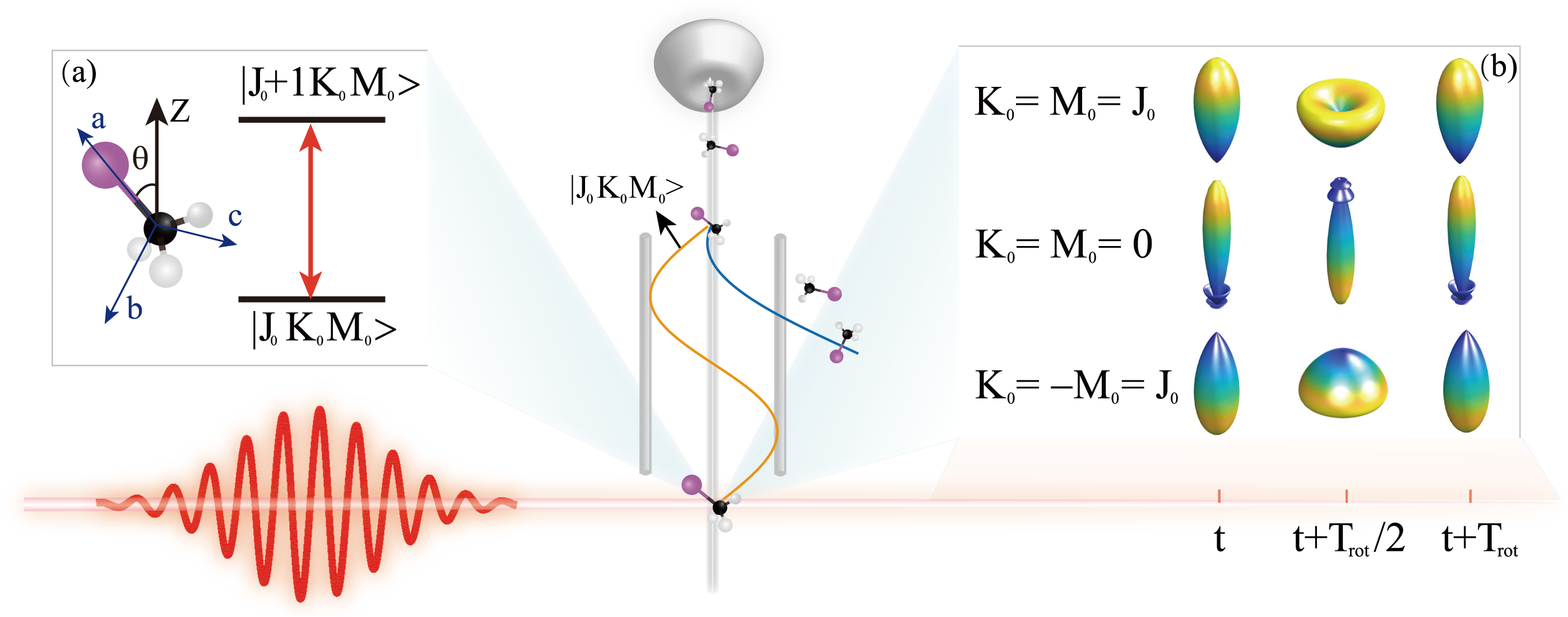}}\caption{Schematic illustration of single-pulse control method for enhancing unidirectional field-free orientation in symmetric molecules.
The central panel illustrates the theoretical concept of the control process: an electrostatic hexapole selector prepares and focuses a state-selected molecular ensemble in the initial rotational state \( |J_0K_0M_0\rangle \), which is subsequently driven by a single resonant control pulse that couples this state to the adjacent rotational state \( |J_0+1K_0M_0\rangle \). Panel (a) shows the corresponding two-state excitation model, where \( \theta \) denotes the angle between the molecular $a$ axis and the polarization direction of the laser pulse, \( J_0 \), \( K_0 \) and \( M_0 \) represent the total angular momentum and its projections in the molecule-fixed $z$ and space-fixed $Z$ axes, respectively. Panel (b) presents the angular distributions of the rotational wave packet after the pulse, showing the periodic evolution of orientation and its dependence on the initial state.}
\label{fig1}
\end{figure*}
\section{Theoretical Methods} \label{TM}
\subsection{Single-pulse control of symmetric top molecules}
Figure \ref{fig1} demonstrates the single-pulse control scheme, which is designed to manipulate two rotational quantum states of symmetric-top molecules, starting from a selected initial state. For illustration, we focus on symmetric-top molecules in the prolate limit ($A > B = C$), where $A$, $B$, and $C$ represent the rotational constants associated with the principal axes $a$, $b$, and $c$ of the molecule, as shown in Fig. \ref{fig1}(a). By assigning the molecule-fixed axes as $a\to z$, $b\to x$, and $c\to y$, the field-free Hamiltonian can be expressed (with $\hbar=1$) as $\hat{H}_{0}=\hat{H}_s+\hat{H}_d$. Here, the first term, $\hat{H}_s = C\hat{J}^2+(A-C)\hat{J}_z^2$, describes the rigid rotor using the angular momentum operator $\hat{J}$ and its projection $\hat{J}_z$ along the $z$ axis, while the second term, $\hat{H}_d = -D_J\hat{J}^4 - D_{JK}\hat{J}^2\hat{J}_z^2 - D_K\hat{J}_z^4$, accounts for centrifugal distortion with the centrifugal constants $D_J$, $D_{JK}$, and $D_{K}$ \cite{2016PRA_Sugny}. Consequently, the eigenenergy \( E_{JK} \) of $\hat{H}_0$, corresponding to the eigenstate \( |JKM\rangle \), comprises the rigid rotor energy \( E^{(r)}_{JK} = C J(J+1) + (A-C) K^2 \) and the centrifugal distortion energy \( E^{(d)}_{JK} = -D_J J^2 (J+1)^2 - D_{JK} J(J+1) K^2 - D_K K^4 \). $J$ denotes the quantum number of the total angular momentum, while the quantum numbers \( M \) and \( K \) (with \( M, K = -J, -J+1, \ldots, J \)) characterize the rotation about the space-fixed \( Z \) axis and the molecule-fixed \( z \) axis, respectively. \\ \indent
We apply a single linearly polarized control pulse at the initial time $t_0$ to the molecular sample that has been selectively prepared in an initial rotational state \(\left | J_0K_0M_0 \right \rangle\) from a thermally populated ensemble, as illustrated in Fig. \ref{fig1}(a). The corresponding control Hamiltonian is expressed as follows
\begin{equation} \label{CH}
\begin{aligned}
    \hat{H}_c(t)=&-\mu_0\mathcal{E}(t)\sum_{J=J_i}\mathcal{M}^{(K_0M_0)}_{J,J}\left|JK_0M_0\right\rangle\left\langle JK_0M_0\right|\\
    &-\mu_0\mathcal{E}(t)\sum_{J=J_i}\mathcal{M}^{(K_0M_0)}_{J,J\pm1}(\left|JK_0M_0\right\rangle\left\langle J\pm1K_0M_0\right|\\
    &+\left|J\pm1K_0M_0\right\rangle\left\langle JK_0M_0\right|),
\end{aligned}
\end{equation}
where $J_i=|K_0|$, $\mu_0$ is the permanent dipole moment of the molecule, and the transition matrix element is defined as $\mathcal{M}^{(K_0M_0)}_{J,J'}=\left\langle JK_0M_0\right|D_{00}^1 \left | J'K_0M_0 \right \rangle$, with $D_{00}^1$ being the element of the Wigner $D$-matrix. In symmetric top molecules, the diagonal elements of $\mathcal{M}^{(K_0M_0)}_{J,J}=M_0K_0/J(J+1)$ can be non-zero with $K_0M_0\ne0$ \cite{1988Zare}. For diatomic and linear symmetric molecules, however, these elements are zero, which leads to the exclusion of the first term on the right-hand side in Eq. (\ref{CH}). The time-dependent rotational wave function $\left|\psi(t)\right\rangle$ is obtained by numerically solving Schrödinger equation under the full Hamiltonian $\hat{H}(t)=\hat{H}_0+\hat{H}_c(t)$. The degree of molecular orientation is then evaluated as $\langle \psi(t) | \cos\theta | \psi(t) \rangle$.
\subsection{Unidirectional field-free orientation in two-state model}
We aim to design the control pulse to create a target superposition comprising two rotational states $|J_0K_0M_0\rangle$ and $|J_0+1K_0M_0\rangle$
\begin{equation}\label{psiT}
\begin{aligned}
    \left | \psi_{\rm{target}}(t)\right \rangle&=c_{J_0K_0M_0}(t)e^{i\varphi_{J_0K_0M_0}}e^{-iE_{J_0K_0}t}\left | J_0K_0M_0\right \rangle\\&+c_{J_0+1K_0M_0}(t)e^{i\varphi_{J_0+1K_0M_0}}e^{-iE_{J_0+1K_0}t}\left | J_0+1K_0M_0\right \rangle,
\end{aligned}
\end{equation}
where $c_{JK_0M_0}$ and $\varphi_{JK_0M_0}$ denote the real positive coefficient and phase of the rotational state.
The corresponding time-dependent degree of molecular orientation can be quantified by
\begin{align}\label{cos}
       \left \langle \cos\theta \right \rangle(t)&=c^2_{J_0K_0M_0}(t)\mathcal{M}^{(K_0M_0)}_{J_0,J_0}+c^2_{J_0+1K_0M_0}(t)\mathcal{M}^{(K_0M_0)}_{J_0+1,J_0+1}\\ \nonumber
       &+2c_{J_0+1K_0M_0}(t)c_{J_0K_0M_0}(t)\mathcal{M}^{(K_0M_0)}_{J_0+1,J_0}\\&\times\cos(\omega^{(K_0)}_{J_0+1,J_0}t-\varphi_{J_0+1,J_0}),
\end{align}
where $\omega^{(K_0)}_{J_0+1,J_0}=E_{J_0+1K_0}-E_{J_0K_0}$ denotes the transition frequency and $\varphi_{J_0+1,J_0}=\varphi_{J_0+1K_0M_0}-\varphi_{J_0K_0M_0}$ the relative phases between states $|J_0K_0M_0\rangle$ and $|J_0+1K_0M_0\rangle$. The transition matrix elements take the values of $\mathcal{M}^{(K_0M_0)}_{J_0,J_0}=M_0K_0/J_0(J_0+1)$, $\mathcal{M}^{(K_0M_0)}_{J_0+1,J_0+1}=M_0K_0/(J_0+1)(J_0+2)$ and $\mathcal{M}^{(K_0M_0)}_{J_0+1,J_0}=\sqrt{\Big[(J_0+1)^2-K_0^2\Big]\Big[(J_0+1)^2-M_0^2\Big]}/(J_0+1)\sqrt{(2J_0+1)(2J_0+3)}$ \cite{2009JCP_Stapelfeldt}. When the molecule is initially in rotational states with $K_0=M_0=0$, the diagonal matrix element $\mathcal{M}^{(K_0M_0)}_{J,J}$ becomes zero. The corresponding time-dependent degree of orientation for symmetric-top molecules resembles that of diatomic and linear polar molecules, exhibiting periodic changes in orientation direction.\\ \indent
By analyzing Eq. (\ref{cos}), we note that when \( J_0 > 0 \) and the values of \( K_0 \) and \( M_0 \) meet specific conditions (namely \( K_0M_0=\pm J_0^2\)), the first two terms can be controlled to consistently yield positive or negative maximum values. When the absolute value of their sum exceeds that of the third term, the time-dependent degree of orientation, \(\langle\cos\theta\rangle(t)\), can consistently be either positive or negative, illustrating the phenomenon of unidirectional field-free orientation in Fig. \ref{fig1}(b). It implies that the maximum orientation can be enhanced by manipulating the population distribution of  two rotational states. In this work, we assume molecules are initially prepared in a well-defined rotational state rather than a degenerate level. This allows for quantitative analysis of how the initial state influences the maximum achievable orientation. State-selected molecules with degeneracy share the same sign of $K_0M_0$, resulting in constructive contributions to orientation, as shown in Eq. (\ref{cos}).\\ \indent
By employing the method of Lagrange multipliers \cite{2020PRA_Niels}, we can derive that $\left \langle \cos\theta \right \rangle(t)$ at full revivals has two extreme values (see Appendix A)
\begin{equation}\label{lm}
\begin{aligned}
\lambda_\pm&=\frac{\mathcal{M}^{(K_0M_0)}_{J_0,J_0}+\mathcal{M}^{(K_0M_0)}_{J_0+1,J_0+1}}{2} \\&\pm \frac{\sqrt{(\mathcal{M}^{(K_0M_0)}_{J_0,J_0}-\mathcal{M}^{(K_0M_0)}_{J_0+1,J_0+1})^2+(2{\mathcal{M}^{(K_0M_0)}_{J_0+1,J_0}})^2}}{2}.
\end{aligned}
\end{equation}
The corresponding amplitudes of the two rotational states required to achieve the extreme orientation are given by
\begin{equation}\label{cjm}
\begin{aligned}
    c_{J_0K_0M_0}=\frac{\mathcal{M}^{(K_0M_0)}_{J_0+1,J_0}}{\sqrt{(\lambda-\mathcal{M}^{(K_0M_0)}_{J_0,J_0})^2+(\mathcal{M}^{(K_0M_0)}_{J_0+1,J_0})^2}},\\ \quad c_{J_0+1K_0M_0}=\frac{\lambda-\mathcal{M}^{(K_0M_0)}_{J_0,J_0}}{\sqrt{(\lambda-\mathcal{M}^{(K_0M_0)}_{J_0,J_0})^2+(\mathcal{M}^{(K_0M_0)}_{J_0+1,J_0})^2}}.
\end{aligned}
\end{equation}
Both extreme values of orientation associated with the coefficients of rotational states are significantly dependent on the values of $K_0$ and $M_0$, as indicated by Eqs. (\ref{lm}) and (\ref{cjm}). Thus, the generation of coherent superposition can enhance molecular orientation more effectively than merely applying inhomogeneous electric fields, which orient molecules of a specific initial state $| J_0K_0M_0\rangle$
\subsection{Pulse-area theorem for control-field design}
We now derive a two-state pulse-area theorem to design the single control pulse for generating the desired rotational superposition in Eq. (\ref{psiT}) with the optimal amplitudes given by Eq. (\ref{cjm}). By employing  the first-order Magnus approximation to describe the unitary operator of the two-state model,  we can derive an analytical wavefunction of the system as \cite{2022PRA_Shu}
\begin{equation}\label{psi1}
    \left | \psi^{(1)}(t)\right \rangle=\cos\Theta (t)\left | J_0K_0M_0\right \rangle+i\sin\Theta(t) e^{-i\phi}\left | J_0+1K_0M_0\right \rangle,
\end{equation}
where the pulse-area $\Theta(t)$ reads   $\Theta(t)=\left|\mu^{(K_0M_0)}_{J_0+1,J_0}\int_{t_0}^{t}\mathcal{E}(t')e^{-i\omega^{(K_0)}_{J_0+1,J_0}t'}dt'\right|$ with its transition dipole moment $\mu^{(K_0M_0)}_{J_0+1,J_0}=\mu_0\mathcal{M}^{(K_0M_0)}_{J_0+1,J_0}$, and $\phi$ corresponds to the phase of the pulse. By comparing the coefficients of the analytical wavefunction in Eq. (\ref{psi1}) with those of the target wavefunction in Eq. (\ref{psiT}), we can determine the extreme values of the orientation in Eq. (\ref{lm}), provided that the control pulse at the final time \(t_f\)  satisfies the pulse-area condition
\begin{equation}\label{area}
    \Theta(t_f)=\arccos( c_{J_0K_0M_0}).
\end{equation}
\begin{figure}[ht]
\centering
\resizebox{0.45\textwidth}{!}{%
\includegraphics{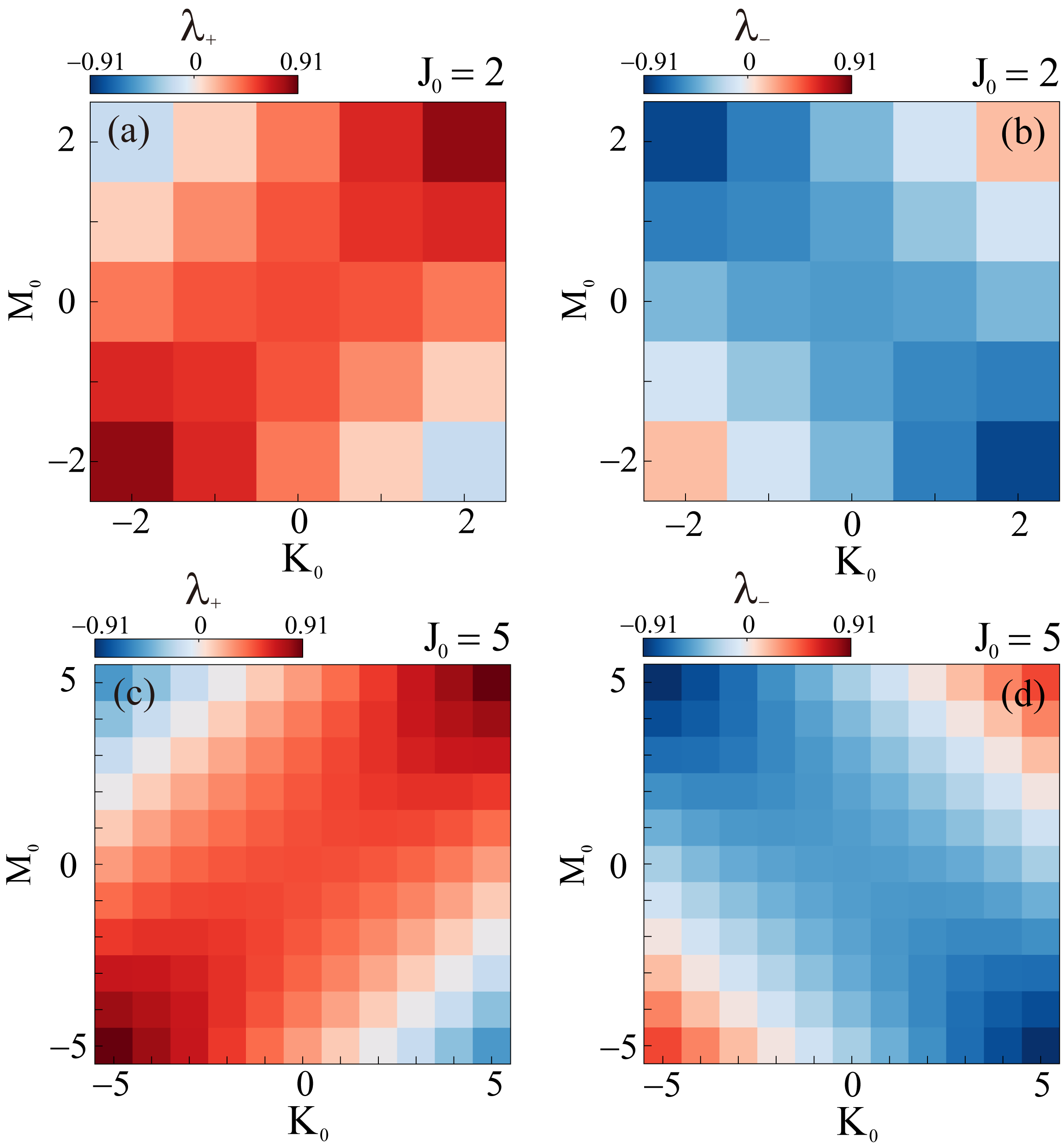}}\caption{Orientation extremes $\lambda_\pm$ versus ($K_0$, $M_0$) for:
(a,b) $J_0=2$ and (c,d) $J_0=5$ manifolds. }   \label{fig2}
\end{figure}\\
We utilize the pulse-area condition in Eq. (\ref{area}) to design the control (terahertz or microwave) pulse, expressed by $\mathcal{E}(t)=\mathcal{E}_0 f(t)\cos[\omega_0 t+\phi]$ with the electric field strength $\mathcal{E}_0$, the center frequency $\omega_0$ and the profile function $f(t)$. Considering a resonant excitation at \(\omega_0 = \omega^{(K_0)}_{J_0 + 1, J_0}\), the pulse area simplifies to \(\Theta(t_f) =\frac{1}{2} \mu^{(K_0M_0)}_{J_0 + 1, J_0} \mathcal{E}_0 \int_{t_0}^{t_f} f(t) dt \). This indicates that the temporal shape of the control pulse can be arbitrary, provided that its profile satisfies the condition: \(\mathcal{E}_0 \int_{t_0}^{t_f} f(t) dt =2\arccos(c_{J_0 K_0 M_0})/\mu^{(K_0M_0)}_{J_0 + 1, J_0}\). For illustration, we employ a control pulse (terahertz or microwave) with a Gaussian profile in the following simulations. The corresponding time-dependent electric field can be expressed as follows
\begin{equation}\label{Et}
    \mathcal{E}(t)=\sqrt{\frac{2}{\pi}}\frac{\Theta(t_f)}{\mu^{(K_0M_0)}_{J_0+1,J_0}\tau}e^{-\frac{t^2}{2\tau^2}}\cos(\omega_0 t+\phi),
\end{equation}
where $\tau$ is the pulse duration. For practical applications, the optimized single pulse in Eq. (\ref{Et}) can be identified by adjusting the pulse strength and its central frequency while measuring the maximum degree of orientation.
\section{Results and Discussion} \label{RD}
\subsection{Dependence of orientation extrema on the initial states}
To validate the theoretical analysis presented in Eqs. (4) and (5), we first examine the dependence of the two extreme orientation values, $\lambda_{+}$ and $\lambda_{-}$, on the initial quantum numbers $K_0$ and $M_0$. Figure \ref{fig2} shows the results for the \( J_0 = 2 \) and \( J_0 = 5 \) manifolds. For the \( J_0 = 2 \) case, the maximum positive orientation value, \( \lambda_{+} = 0.827 \), is achieved at \( K_0 = M_0 = \pm2 \). In contrast, the negative maximum value, \( \lambda_{-} = -0.827 \), occurs at \( K_0 = -M_0 =\pm2 \), as shown in Figs. \ref{fig2}(a) and (b). A similar pattern is observed for the case \( J_0 = 5 \) in Figs. \ref{fig2} (c) and (d), where the positive maximum \( \lambda_{+} = 0.908 \) is found in \( K_0 = M_0 =\pm5 \), while the negative maximum \( \lambda_{-} = -0.908 \) occurs at \( K_0 = -M_0 =\pm5 \). The results illustrated in Fig. \ref{fig2}, along with additional simulations for various values of \( J_0 \), indicate that the maximum positive orientation satisfies the condition \( K_0 M_0 = J_0^2 \) with \( K_0 = M_0 \), whereas the maximum negative orientation requires \( K_0 M_0 = -J_0^2 \) with \( K_0 = -M_0 \), fully aligning with the theoretical analysis presented in Eq. (3).
\begin{figure}[b]
\centering
\resizebox{0.48\textwidth}{!}{%
\includegraphics{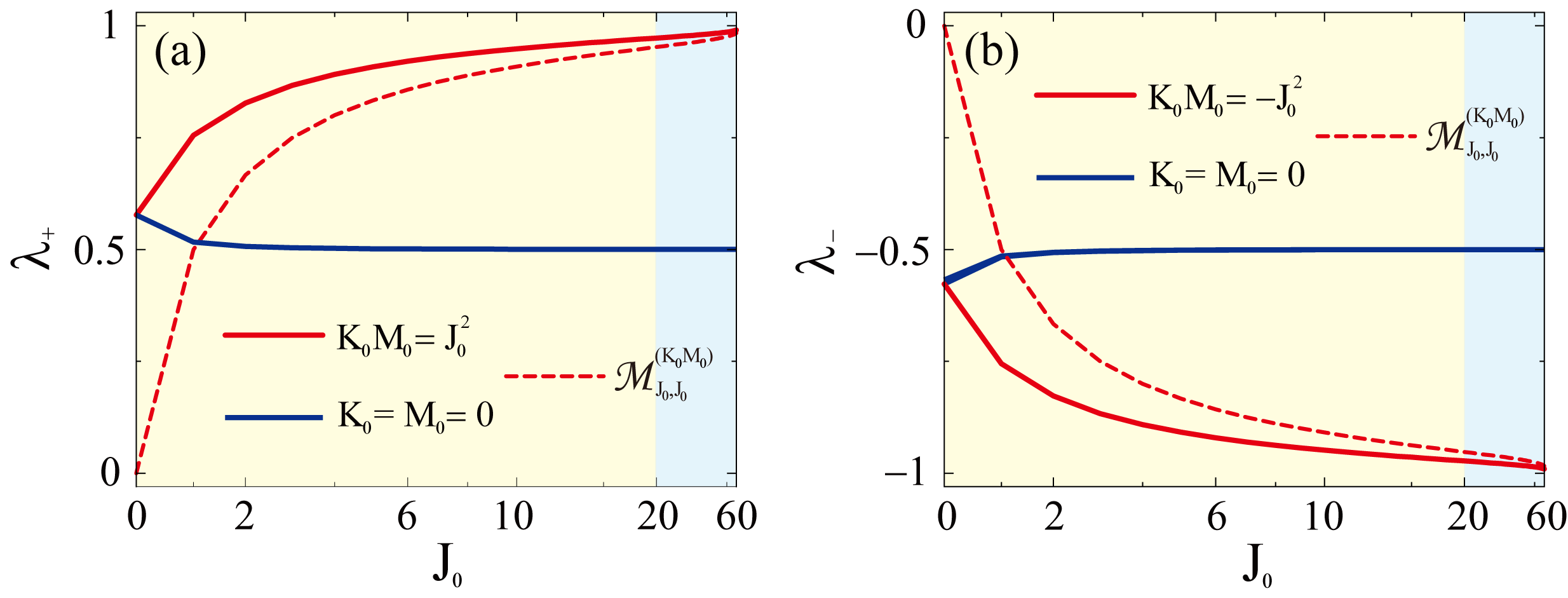}}\caption{The dependence of the two extremes $\lambda_\pm$ of orientation on the discrete values of the initial rotational quantum number $J_0$: (a) $\lambda_+$ for $K_0=M_0=0$ (blue line) and $K_0M_0=J_0^2$ (red line), and (b) $\lambda_-$ for $K_0=M_0=0$ (blue line) and $K_0M_0=-J_0^2$ (red line). The values of  \(\mathcal{M}_{J_0J_0}^{(K_0M_0)}\) (red dashed lines) correspond to the results using pure inhomogeneous electric fields. }
\label{fig3}
\end{figure}
\begin{figure*}[t]
\centering
\resizebox{0.9\textwidth}{!}{%
\includegraphics{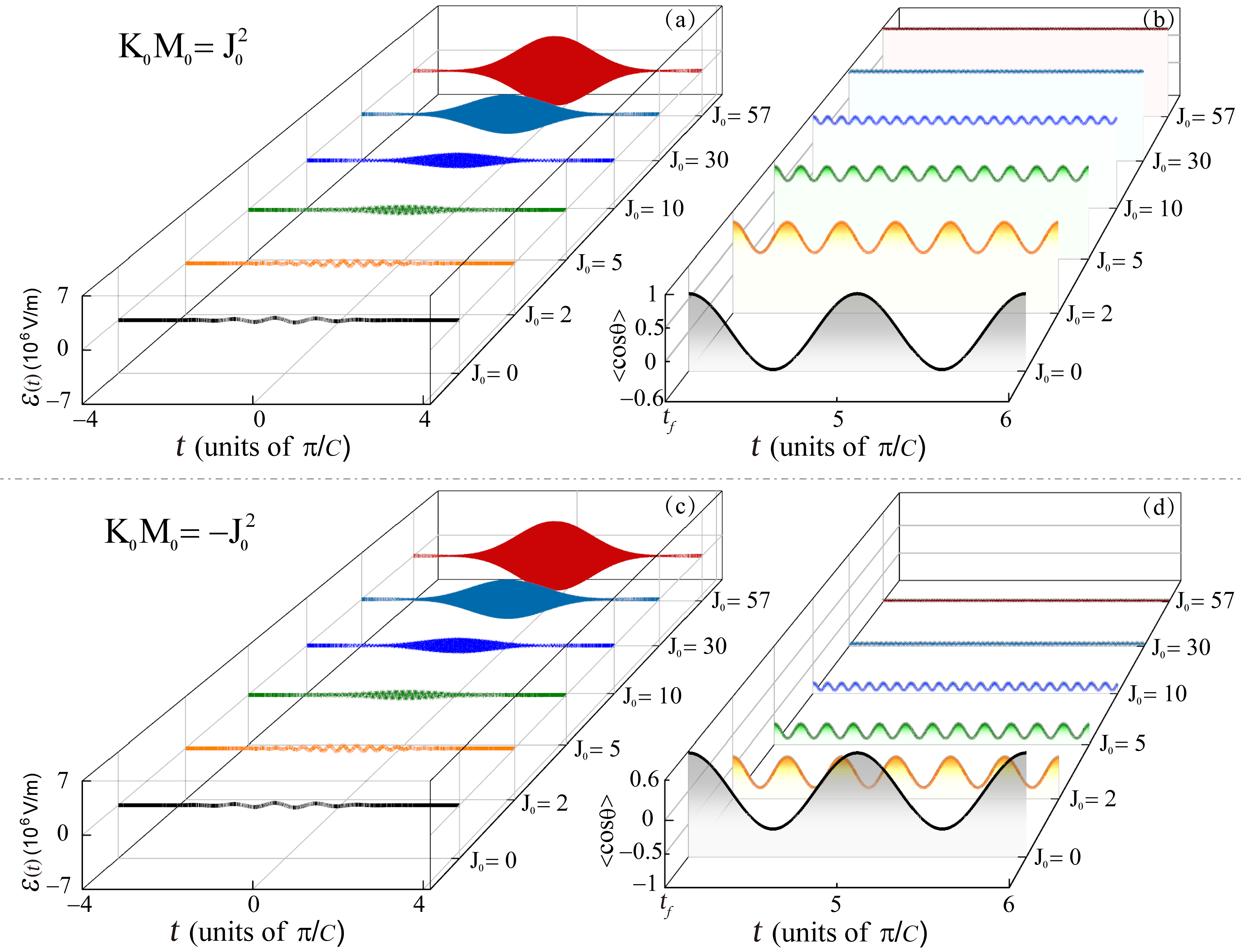}}\caption{Microwave pulse design and orientation dynamics: (a,c) Analytically optimized pulse waveforms (Eq. (\ref{Et})) for different rotational states with (a) \( K_0M_0 = J_0^2 \) and (c) \( K_0M_0 = -J_0^2 \). Corresponding orientation dynamics \(\langle \cos\theta \rangle\) shown in (b,d), demonstrating \( J_0 \)-dependent control efficiency.}
\label{fig4}
\end{figure*}
Following the verification that the extreme values of orientation occur at \( K_0M_0=\pm J_0^2 \) for a specific \( J_0 \) manifold, we proceed to analyze the effect of the initial rotational quantum number \( J_0 \) on the orientation extrema.
 Figure \ref{fig3} illustrates the variation of the extreme orientation values, \(\lambda_+\) and \(\lambda_-\), with \(J_0\). For the initial conditions where \(K_0 = M_0 = 0\), coupling the initial state \(J_0 = 0\) with the excited state \(J = 1\) results in the maximal  orientation value of \(|\langle \cos\theta \rangle|_{\rm{max}} =0.577\). As \(J_0\) increases, the maximum orientation value decreases rapidly to its limit of $\pm0.5$, which is the same as that observed in diatomic and linear symmetric molecules. When the quantum numbers \(K_0\) and \(M_0\) satisfy the conditions \(K_0M_0=\pm J_0^2\), we observe that increasing \(J_0\) yields can significantly enhance positive and negative orientation values. For $J_0=5$, the maximum orientation can be increased to high values of $\pm0.908$ with $K_0=\pm M_0=\pm5$. Specifically, the maximum orientation \(|\langle \cos\theta \rangle|_{\rm{max}}\) can exceed 0.99 for molecules initially in the state of \(J_0 = 57\), which greatly simplifies the complexity of the required pulses and the involved rotational states compared to ultracold diatomic molecules \cite{2025PRR_Shu}. This highlights the significant roles of the selected initial states (with specific quantum numbers \(K_0\) and \(M_0\)) and their coherent excitation to neighboring rotational states in enhancing the maximum orientation of symmetric top molecules. This behavior essentially differs from the observations shown in diatomic and linear symmetric molecules.

To illustrate how rotational excitation enhances maximum orientation, Fig. \ref{fig3} also presents the diagonal matrix elements $\mathcal{M}_{J_0, J_0}^{K_0M_0}$ as a function of the rotational quantum number $J_0$, corresponding to an approach that employs only inhomogeneous electric fields. The rotational excitation of specific initial states leads to significant enhancements in molecular maximum orientation, particularly for values of \(J_0 < 20\). We note that while selecting extreme initial rotational states  (e.g., $J_0>60$) can theoretically lead to considerable unidirectional molecular orientation, the generation of the corresponding inhomogeneous electric fields poses significant experimental challenges. This complexity arises from the need to consider both first-order and second-order Stark effects simultaneously.
\subsection{Numerical simulations for symmetric-top molecules}
We now conduct simulations by numerically solving the time-dependent Schrödinger equation to examine the theoretical maximum orientation of methyl iodide (\(\mathrm{CH}_3\mathrm{I}\)) induced by the analytically designed control pulse in Eq. (\ref{Et}). The molecular parameters used in the simulations are \(A = 5.173949 \, \mathrm{cm}^{-1}\), \(C = 0.25098 \, \mathrm{cm}^{-1}\), \(D_J = 2.1040012 \times 10^{-7} \, \mathrm{cm}^{-1}\), \(D_{JK} = 3.2944780 \times 10^{-6} \, \mathrm{cm}^{-1}\), \(D_{K} = 8.7632195 \times 10^{-5} \, \mathrm{cm}^{-1}\), and \(\mu_0 = 1.6406 \, \mathrm{D}\) \cite{2016PRA_Sugny}. To eliminate the dc component, we set the phase of the pulses to \(\phi = \frac{\pi}{2}\).
The amplitude of the pulse meets the criteria specified in Eq. (\ref{area}), with the pulse duration set at \(\tau = \pi/C\). This duration aligns with the rotational revival period \(T_{\mathrm{rot}} = 2\pi/\omega_{1,0}^{(0)}\) for molecules initially in the rotational ground state and  is sufficiently long to prevent the excitation of additional rotational states by limiting the pulse    bandwidth \cite{sm}.\\ \indent
Figure \ref{fig4} presents the analytically designed microwave pulses \(\mathcal{E}(t)\) alongside the corresponding time-dependent degrees of orientation \(\left \langle \cos\theta \right \rangle(t)\) after the pulses are turned off, for six different initial states \(J_0=0, 2, 5, 10, 30, 57\). The parameters \(K_0\) and \(M_0\) satisfy the condition \(K_0M_0=\pm J_0^2\). For molecules initially in the \(J_0=0\) state, the degree of orientation undergoes periodic evolution over time, exhibiting the maximum positive and negative orientations of \(\pm 0.577\) at full revivals, as shown in Figs. \ref{fig4}(b) and (d). These findings align with those observed in diatomic and linear symmetric molecules. As the initial quantum number \(J_0\) increases, the oscillations in orientation become increasingly rapid due to the rising rotational transition frequency \(\omega_{J_0+1, J_0}^{(K_0)}\) with \(J_0\), resulting in shorter rotational periods. Interestingly, for the initial state \( J_0 > 0 \), the time-dependent orientation degree displays periodic oscillations, with values that consistently turn positive in Fig. \ref{fig4}(b) and negative in Fig. \ref{fig4}(d). This phenomenon signifies the occurrence of unidirectional field-free orientation, and its maximum value increases as \( J_0 \) rises.\\ \indent
\begin{figure}[ht]
\centering
\resizebox{0.48\textwidth}{!}{%
\includegraphics{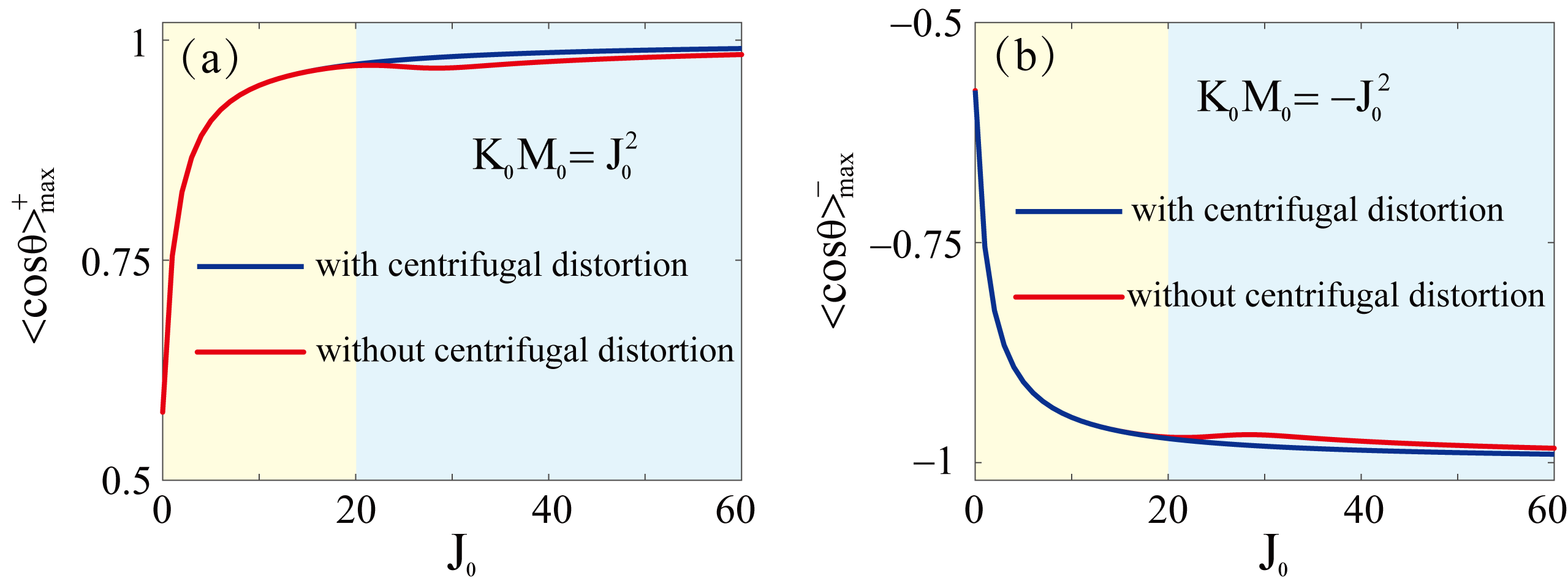}}\caption{Orientation maxima comparison: Full model (blue line) vs. centrifugal-distortion-excluded model (red line) for different discrete values of the initial rotational quantum number \( J_0 \).  Results are shown for (a) \( K_0M_0 = J_0^2 \) and (b) \( K_0M_0 = -J_0^2 \).  The phases and durations of the microwave  pulses used for calculations are the same as those in Fig. \ref{fig4}.}
\label{fig5}
\end{figure}
As shown in Fig. \ref{fig4}(b), the maximum positive value exceeds 0.99 for the initial state with $J_0=57$, $M_0=K_0=\pm57$, while in Fig. \ref{fig4}(d), the maximum negative value drops below -0.99 for the initial state with $J_0=57$, $M_0=-K_0=\pm57$, corresponding to the near-perfect unidirectional orientation. To gain insights into the underlying physics, we analyze the final population distribution, which reveals that 0.85 of the population resides in the state \(J = 57\), while the remaining 0.15 is at \(J = 58\). Such a population distribution can optimize the third term in Eq. (\ref{cos}), although it declines sharply with increasing \(J_0\) and becomes significantly reduced at \(J_0 = 57\). This results in a rapid decrease of the oscillation amplitude in the time-dependent orientation degree. Achieving the same maximum orientation without the specially designed pulse would need the molecule in an extreme rotational state with \(J_0\ge99 \), as shown by the exclusion of the second and third terms in Eq. (\ref{cos}).  This demonstrates that our method can reduce the need for an extreme initial rotational state, facilitating faster convergence towards high orientation values above 0.99, which are essential for practical applications in precise spectroscopy and quantum information techniques.
\\ \indent
We now examine the impact of centrifugal distortion on the application of our methodology.  Figure \ref{fig5} depicts the positive and negative orientation maxima (\(\left \langle \cos\theta \right \rangle_{\mathrm{max}}^{\pm}\)) calculated using the complete Hamiltonian \(\hat{H}_0\) compared with the results that exclude the centrifugal distortion Hamiltonian \(\hat{H}_d\). For this comparison,  we apply identical analytically designed pulses, based on Eq. (\ref{Et}), to both models for the corresponding initial quantum number \(J_0\).  We can see that the discrepancies for initial quantum numbers \(J_0<20\) are negligible, but they become noticeable when the molecules are initially in higher rotational states with \(J_0>20\). This indicates that including the centrifugal distortion term in the model is essential for designing the pulse that achieves nearly perfect unidirectional orientation. For the sake of experimental convenience in designing inhomogeneous electric fields, the selected initial quantum number \( J_0 \) can be set to less than 20, which helps avoid centrifugal distortion effects while enhancing quantum coherence.\\ \indent
To evaluate the reliance of the proposed scheme on key pulse parameters, we introduce perturbations to both the central frequency and amplitude of the designed pulse. Specifically, the central frequency and pulse area are defined as
\begin{equation}
\omega_0 = (1+\varepsilon_1)\omega_{J_0+1,J_0}^{(K_0)} \end{equation}and
\begin{equation}\theta(t_f) = (1+\varepsilon_2)\arccos{(c_{J_0K_0M_0})},
\end{equation}
where \(\varepsilon_1\) and \(\varepsilon_2\) denote the relative deviations in frequency and amplitude. To quantify the effects of these deviations on the enhancement efficiency, we introduce an enhancement fidelity  \(\eta\), defined as
\begin{equation}
\eta = 1 - \frac{\lambda_{\pm} - \langle \cos\theta \rangle^{\pm}}{\lambda_{\pm} - \mathcal{M}^{(K_0M_0)}_{J_0,J_0}}.
\end{equation}
\(\eta = 1\) indicates the maximum enhancement, where the designed pulse aligns with the theoretical limit, while \(\eta = 0\) suggests no enhancement, leaving the molecule in the selected initial state.\\ \indent
\begin{figure*}[ht]
\centering
\resizebox{1\textwidth}{!}{%
\includegraphics{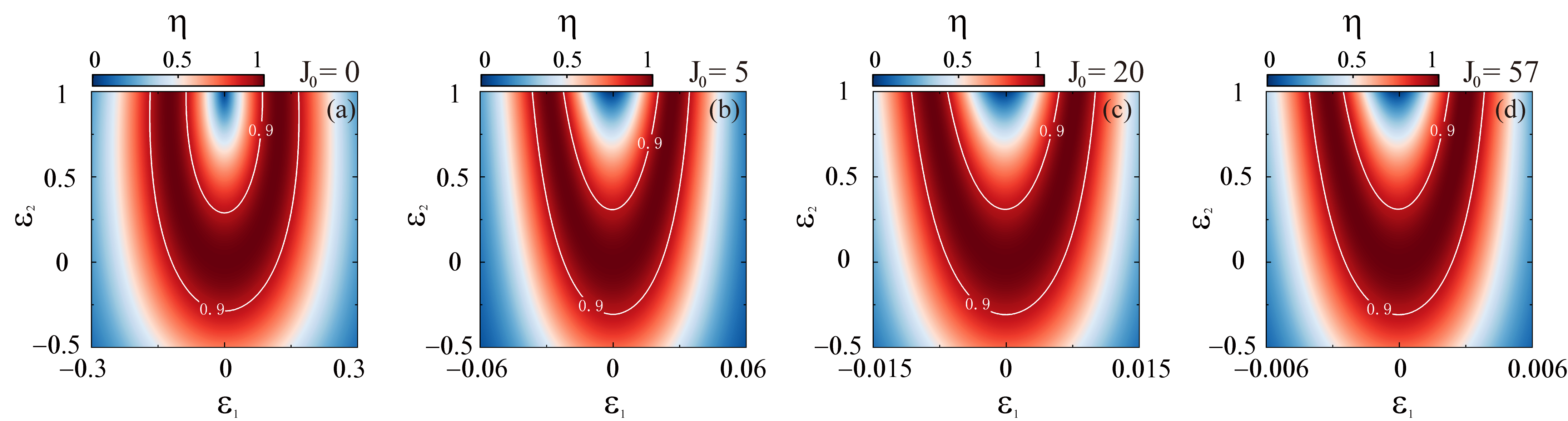}}\caption{Efficiency $\eta$ of generating the maximal orientation as a function of relative fluctuations in center frequency ($\varepsilon_1$) and pulse amplitude ($\varepsilon_2$), at fixed pulse duration $\tau=\tau_0$. Panels (a)–(d) correspond to initial states $J_0=0$, 5, 20, and 57 for $K_0M_0=J_0^2$.}
\label{S2}
\end{figure*}
Figure \ref{S2} presents \(\eta\) as a function of \(\varepsilon_1\) and \(\varepsilon_2\) for initial states with \(J_0 = 0, 5, 20, 57\), using \(K_0M_0 = J_0^2\) as an example. The other simulation parameters are consistent with those used in Fig. 3 of the main text. For the ground state (\(J_0 = 0\)), the enhancement efficiency shown in Fig. \ref{S2}(a) exceeds \(90\%\) even with frequency deviations of up to \(\pm15\%\) and amplitude deviations of up to \(\pm25\%\), demonstrating the robustness of the scheme. In higher initial states (\(J_0 > 0\)), as depicted in Figs. \ref{S2}(b)-(d), the efficiency is less sensitive to amplitude fluctuations but becomes sensitive to variations in central frequency. For \(J_0 = 20\), maintaining the enhancement above \(90\%\) requires limiting frequency deviation to below \(1.0\%\). For \(J_0 = 57\), frequency deviation should be constrained within \(0.1\%\). It is essential to note that for higher initial rotational states, the pulse duration can be shortened, as illustrated in Fig. \ref{S2}, which can reduce the scheme's sensitivity to fluctuations in the central frequency $\omega_0$. This robustness to key parameter deviations could further reduce experimental complexity by generating microwave pulses with the necessary precision and stability.
\section{Conclusion and Outlooks}\label{CO}
We established a theoretical framework for generating and enhancing the unidirectional field-free orientation of symmetric top molecules, starting from a selected rotational state \( |J_0K_0M_0\rangle \). Our analysis indicates that creating an optimal superposition of two rotational states, under the condition \( K_0 M_0 = \pm J_0^2 \), can significantly enhance the maximum orientation of molecules that are initially prepared in a considerably lower rotational state.  By deriving a two-state pulse-area theorem, we developed a quantum control strategy to obtain the desired unidirectional molecular orientation through the analytical design of a single control pulse. This finding highlights the crucial roles of initial-state selection and coherent superpositions in attaining stable, unidirectional orientation, thereby providing a theoretical foundation for the experimental realization of high and long-lasting field-free orientation in symmetric top molecules. Numerical simulations demonstrated that applying the analytically designed single pulse to methyl iodide (\(\mathrm{CH}_3\mathrm{I}\)) molecules can effectively enhance unidirectional field-free orientation, surpassing the efficacy of using inhomogeneous electric fields alone. The robustness analysis further demonstrates that the enhancement efficiency remains high within reasonable fluctuations of pulse amplitude and frequency, supporting the practical feasibility of the proposed control scheme. This approach eliminates the need to optimize a superposition of multiple rotational states at ultracold temperatures. It can also be applied to other symmetric top molecules by considering their specific rotational constants and dipole moment parameters.\\ \indent
The experimental realization of this scheme relies on two essential capabilities: (i) preparation of molecules in a well-defined rotational state $\left|J_0K_0M_0\right\rangle$ and (ii) generation of a resonant terahertz or microwave control field. For symmetric-top molecules such as CH$_3$Br and CH$_3$I, electrostatic hexapole focusing has been demonstrated to enrich selected rotational states from a thermal ensemble. Although complete purification remains challenging for high-$J_0$ states, current techniques can yield dominant-state populations sufficient for generating and enhancing unidirectional orientation. Therefore, our assumption of a single-state initial condition should be viewed as an idealized upper limit that defines the maximal orientation efficiency and clarifies experimental requirements. In typical experiments, terahertz pulses have durations on the picosecond timescale, whereas microwave pulses last from hundreds of nanoseconds to microseconds. According to the analytically optimized control field in our scheme, the required terahertz and microwave field amplitudes are on the order of $10^6$ and $10^3$ V/m, respectively, both of which are well within accessible  experimental conditions. Under these conditions, the predicted orientation can be directly observed using  COLTRIMS \cite{2018NC_Wu} and the weak-field polarization technique \cite{2023PRA_shu}. Given recent advances in molecular-state purification \cite{2009NP_Vrakking,2015PRA_Ding,2017JPCA_Ding}, pulse shaping \cite{2021OL_Liu,2023OE_Yu}, and orientation detection \cite{2015SA_Ohshima}, experimental verification of this theoretical scheme, as well as exploration of its potential applications, is highly promising.\\ \indent
Oriented molecules are vital in stereochemistry, spectroscopy, and quantum computing. In stereochemistry, they allow investigation of orientation-dependent reaction dynamics and chiral discrimination. In precision spectroscopy, preparing these ensembles enhances transition selectivity and spectral resolution by minimizing orientation averaging. In quantum computing, control over rotational-state superpositions is key for encoding and manipulating molecular qubits. Unidirectional, field-free orientation of symmetric top molecules could further enhance these advantages by precisely aligning molecular axes without the need for continuous external fields. This reduces perturbations and enables high-fidelity studies of intrinsic molecular properties and reaction pathways. In stereochemistry, it facilitates detailed analysis of stereodynamic effects and chiral selectivity. For spectroscopy, it can nearly eliminate orientation averaging, resulting in superior spectral resolution. In quantum computing, it supports robust initialization and manipulation of molecular qubits. Moreover, initial-state-dependent unidirectional orientation allows non-destructive, weak-field polarization measurements and identification, removing the need for strong fields.\\ \indent

\begin{acknowledgments}
This work was supported by the National Natural Science Foundation of China (NSFC) under Grant No. 12274470 and No. 12134005. D. Dong is supported in part by the Australian Research Council’s Future Fellowship funding scheme under Project FT220100656. Q.-Q. H. acknowledges financial support in part from the Fundamental Research Funds for the Central Universities of Central South University under
Grant No. 1053320231828. The numerical simulation was partially conducted using computing resources at
the High Performance Computing Center of Central South University.
\end{acknowledgments}

\section*{DATA AVAILABILITY}
The data that support the findings of this article are not publicly available.  The data are available from the authors upon reasonable request.

\appendix
\section{Maximum Degree of Orientation within a Two-State Model}
Within the two-state model comprising the rotational states $\left|J_0K_0M_0\right\rangle$ and $\left|J_0+1K_0M_0\right\rangle$, we can determine the extreme values of orientation by employing the method of Lagrange multipliers. The associated Lagrangian functional is defined as
\begin{equation}
    \mathcal{L}(c_{J_0K_0M_0},c_{J_0+1K_0M_0},\lambda)=f-\lambda g,
\end{equation}
with the objective function
\begin{equation}
\begin{aligned}
    f&=c^2_{J_0K_0M_0}\mathcal{M}^{(K_0M_0)}_{J_0,J_0}+c^2_{J_0+1K_0M_0}\mathcal{M}^{(K_0M_0)}_{J_0+1,J_0+1}\\&+2c_{J_0+1K_0M_0}c_{J_0K_0M_0}\mathcal{M}^{(K_0M_0)}_{J_0+1,J_0},
\end{aligned}
\end{equation}
where $\lambda$ represents the Lagrange multiplier constrained by $g=c^2_{J_0K_0M_0}+c^2_{J_0+1K_0M_0}-1=0$. The extremum of $f$, subject to the constraint $g$, can be determined by solving the stationarity equation $\nabla\mathcal{L}=0$, which results in the following relations
\begin{equation}\label{Lm}
    \begin{aligned}
        c_{J_0K_0M_0}\mathcal{M}_{J_0,J_0}^{(K_0M_0)}+ c_{J_0+1K_0M_0}\mathcal{M}_{J_0+1,J_0}^{(K_0M_0)}-\lambda c_{J_0K_0M_0}&=0,\\
        c_{J_0+1K_0M_0}\mathcal{M}_{J_0+1,J_0+1}^{(K_0M_0)}+ c_{J_0K_0M_0}\mathcal{M}_{J_0+1,J_0}^{(K_0M_0)}-\lambda c_{J_0+1K_0M_0}&=0.
    \end{aligned}
\end{equation}
By multiplying the two equations in (\ref{Lm}) by \( c_{J_0K_0M_0} \) and \( c_{J_0+1K_0M_0} \), respectively, and subsequently summing them, we can derive the following relation:
\begin{equation}
    f - \lambda(c_{J_0K_0M_0}^2 + c_{J_0+1K_0M_0}^2) = f - \lambda = 0.
\end{equation}
This relation indicates that the extreme value of the orientation amplitude \( f \) corresponds to the eigenvalue \( \lambda \). Utilizing Eq. (\ref{Lm}), we can also obtain two equivalent expressions for the expansion coefficients
\begin{equation}\label{cm}
    \begin{aligned}
        &c_{J_0+1K_0M_0}=\frac{\lambda-\mathcal{M}_{J_0,J_0}^{(K_0M_0)}}{\mathcal{M}_{J_0+1,J_0}^{(K_0M_0)}}c_{J_0K_0M_0},\\
    &c_{J_0+1K_0M_0}=\frac{\mathcal{M}_{J_0+1,J_0}^{(K_0M_0)}}{\lambda-\mathcal{M}_{J_0+1,J_0+1}^{(K_0M_0)}}c_{J_0K_0M_0}.
    \end{aligned}
\end{equation}
The analysis of Eq. (\ref{cm}) leads to the extremal values of orientation $\lambda_\pm$ and the amplitudes of the two rotational states $c_{J_0K_0M_0}$ and $c_{J_0+1K_0M_0}$ used in Eqs.(\ref{lm}) and (\ref{cjm}).
\begin{figure}[ht]
\centering
\resizebox{0.48\textwidth}{!}{%
\includegraphics{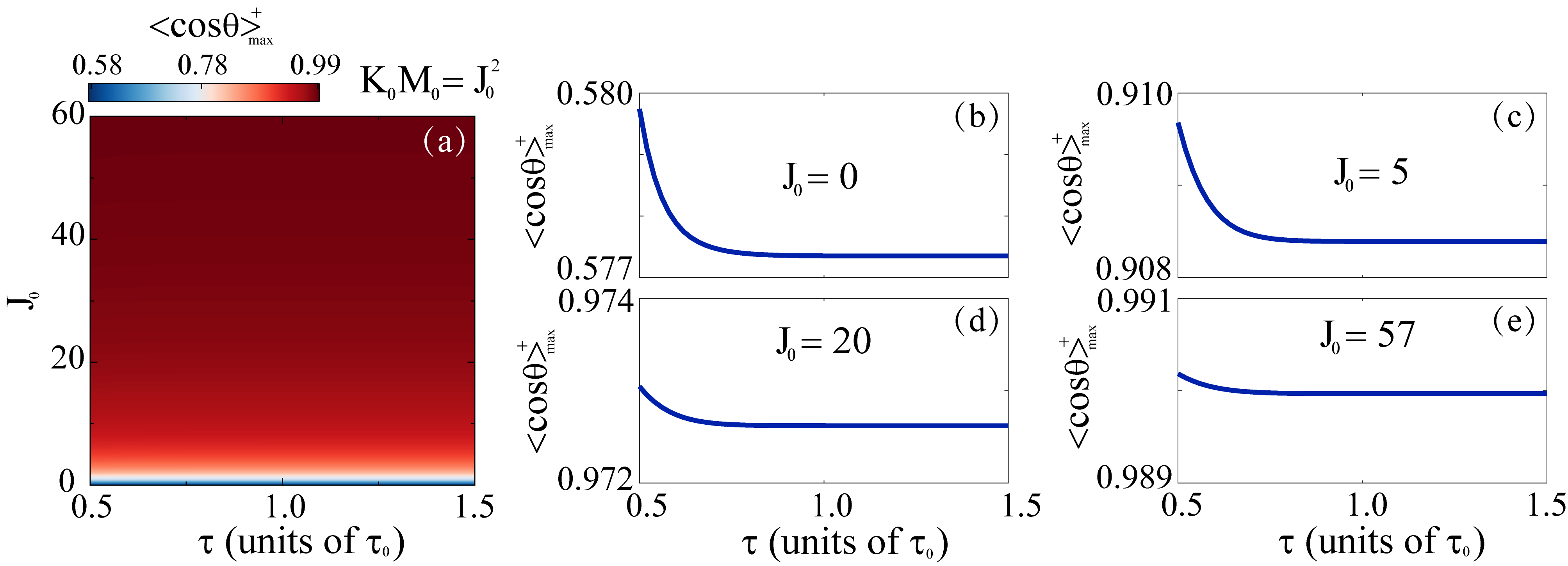}}\caption{Dependence of the maximum positive orientation \(\langle \cos\theta \rangle^{+}_{\text{max}}\) on pulse duration $\tau$ for various initial quantum numbers \(J_0\) for \(K_0M_0 = J_0^2\). The pulse amplitude and phase are the same as those used in Figs. 4 and 5 of the main text. Pulse duration is expressed in units of \(\tau_0 = 2\pi/\omega_{1,0}^{(0)}\).}
\label{S1}
\end{figure}
\section{Dependence of Orientation on Pulse Duration}
To investigate how the proposed method depends on the pulse duration, Fig. \ref{S1}(a) illustrates the maximum positive orientation, \(\langle \cos\theta \rangle^{+}_{\text{max}}\), as a function of the pulse duration \(\tau\), measured in units of \(\tau_0 = 2\pi/\omega_{1,0}^{(0)}\). The simulation examines various initial rotational states, \(J_0\), while adhering to the condition \(K_0 M_0 = J_0^2\). The results show that the pulse duration  affects the maximum degree of orientation. Specifically, the minimum duration required to achieve maximum orientation increases as \(J_0\) rises. To clarify this dependence further, Figs. \ref{S1}(b)-(e) present the maximum degree of orientation for four distinct initial states: \(J_0 = 0\), 5, 20, and 57. For molecules initially in the \(J_0=0\) state, the maximum orientation reaches the theoretical limit of 0.577 at \(\tau = \tau_0\), which corresponds to a two-cycle pulse. When the pulse duration is less than \(\tau_0\) (\(\tau < \tau_0\)), the maximum orientation deviates from this theoretical limit due to the broad bandwidth of the pulse, which leads to unintended excitation of higher rotational states that exceed the two-state excitation model.  As \(J_0\) increases, the minimum pulse duration necessary to achieve maximal orientation decreases. This is because the transition frequency rises with \(J_0\), resulting in a shorter optical period. Consequently, the total optical period involved in the pulse extends for the same duration of the pulse. These findings indicate that the pulse duration of \(\tau = \tau_0\), used in Figs. \ref{fig4}-\ref{S2}, is sufficiently long to selectively excite the target two-state superposition, ultimately leading to maximal orientation.

\bibliography{reference}

\end{document}